# OMAP-L138 LCDK Development Kit

Bharath K P
School of Electronics Engineering
Vellore Institute of Technology
Vellore, India
bharathkp25@gmail.com

Sylash K
School of Electronics Engineering
Vellore Institute of Technology
Vellore, India
sylash.2015@vit.ac.in

Pravina K
School of Electronics Engineering
Vellore Institute of Technology
Vellore, India
pravinakarunanithi@gmail.com

Rajesh Kumar M, *Senior Member, IEEE*
School of Electronics Engineering
Vellore Institute of Technology
Vellore, India
mrajeshkumar@vit.ac.in

*Abstract*— **Low cost and low power consumption processor play a vital role in the field of Digital Signal Processing (DSP). The OMAP-L138 development kit which is low cost, low power consumption, ease and speed, with a wide variety of applications includes Digital signal processing, Image processing and video processing. This paper represents the basic introduction to OMAP-L138 processor and quick procedural steps for real time and non-real time implementations with a set of programs. The real time experiments are based on audio in the applications of audio loopback, delay and echo. Whereas the non-real time experiments are generation of a sine wave, low pass and high pass filter.**

## I. INTRODUCTION

The Texas Instruments (TI) based OMAP-L138 development kit or Low Cost Development Kit (LCDK), have been extensively used in the area of Digital Signal Processing (DSP) in both industry and educational universities [1]. It is a dual core SoC and this processor has features of high geared and low power efficiency. LCDK is a development tool and it is easy to use for both experienced and beginners. Basically users adopts the low cost and low power solutions in several applications in different fields, such as audio, speech communication, biometric, image and video processing.

Since from 2004, the TI promoting the university programs based on the use of TMS320C6713 DSK for both teaching and research. Again TI promotes new LogicPD zoom experimenter in 2011, featuring an OMAP-L138 dual core system on a chip (SoC). This leads to build opportunity to develop C6748 and ARM9 processors. In [2] explains a set of non-real time programming examples and related theory book called Digital Signal Processing and Applications with OMAP-L138 experimenter. Donald S.R [3] describes a set of real time programing examples based on polling method using experimenter processor. The difference between experimenter processor and C6713 DSK with a set of non-real time Programming examples based on polling method is explained [4].

TMS320C6x architecture, audio signal sampling and fixed v/s floating point DSP is described [5]. Rajesh Muthu [6] explains the basic procedural step for non-real time programming in OMAP-L138.

The OMAP-L138 Development kit LCDK[7] having a C674x floating or fixed point DSP with 456MHz performance, it is having On-chip Real Time Clock (RTC), it has DDR2 which is running at 150MHz with NAND flash and the SD card slot. It has serial USB port, VGA and LCD ports. It is having a 3 audio ports line-in and mic-in acts as input ports and line-out acts as output port. It also has an input pin for leopard imaging camera. For OMAP-L138 development kit, Code Composer Studio (CCS) software has been used to develop the experiments and projects. The external connection of a LCDK board is as shown in fig.1.Basically the audio frequency DSP system consists of analog to digital converters and digital to analog converters as well as processor. In this case the LCDK is provided by TMS320C6748 processor. The OMAP-L138 is the part of LCDK processor, which is used for audio, image and video processing. In the applications of an audio the board gives the access through the 3.5mm jack socket for analog inputs and the outputs will be on codec, this can be given to signal generator, digital oscilloscopes, speakers, sound recorders, music players, etc. The OMAP-L138 can be used in DSP applications independently by using the code composer studio which intern connected through JTAG and the XDS100v2 emulator [8] should be used. The XDS100v2 emulator is as shown in fig.2.

In this paper, we are mainly concentrating about the procedural steps for both real time and non-real time implementation and set of real time programs which comes under DSP and non-real time programs based on audio processing. Both real time and non-real time programs are written using interrupt method which is best compare to polling method in case of execution timing. Some the real time experiments like audio loop back, delay in audio, echo and some of the non-real time experiments are sine wave generation, low pass and high pass filters. These programs can be act as a quick reference for the beginners.

In section 2, about the OMAP-L138 development kit and its features are discussed. Procedural steps for non-real time real time implementations and a set of programs with their graphs are plotted and discussed in section 3 and section 4 respectively. Final conclusion and aim for future scope is discussed in section 5.

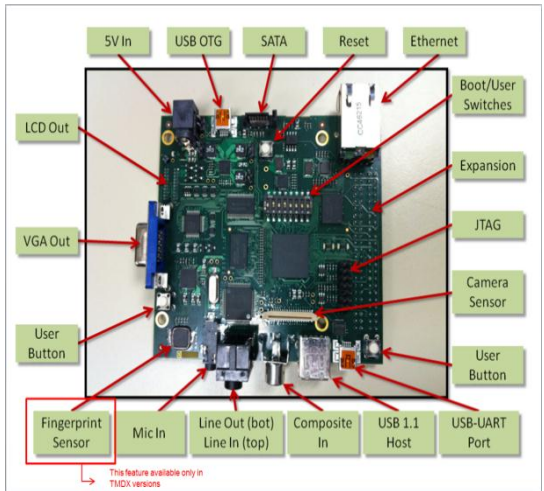

Fig.1. External connection of a LCDK board [7].

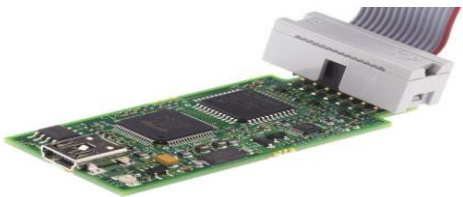

Fig.2 XDS100v2 emulator [8].

## II. OMAP-L138 DEVELOPMENT KIT

The OMAP-L138 development kit (LCDK) is low cost, low power consumption and scalable platform which breaks down advance barriers for the essential applications which requires the embedded system analytics and real-time DSP and also includes the audio, biometric analytics and communications.

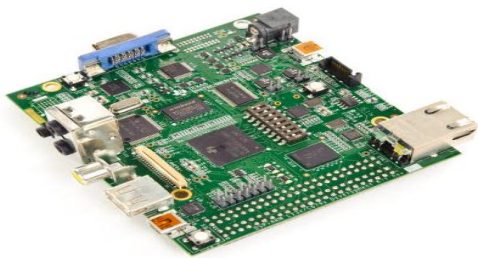

Fig.3 OMAP-L138 development kit (LCDK)

LCDK is also ease and speed the hardware development of real-time signal processing applications. A huge variety of some standard interfaces for connectivity and storage enable that can be easily brings the audio, image and video onto the board. The OMAP-L138 development kit (LCDK) board is as shown in fig.3 and its external connections are already shown in fig.1.

Some of the TMS320C6748 DSP LCDK Features are:
**a.** Highest-Performance Floating-Point Digital Signal Processor (DSP): Eight 32-Bit Instructions/Cycle, 32/64-Bit Data Word, 375/456-MHz C674x Fixed/Floating-Point, Up to 3648/2746 C674x MIPS/MFLOPS, Rich Peripheral Set, Optimized for Audio and Highly Optimized C/C++ Compiler
**b.** Advanced Very Long Instruction Word (VLIW) TMS320C67x™ DSP Core
   ➢ Eight Independent Functional Units:
   ➢ Load-Store Architecture With 64 32-Bit General-Purpose Registers
**c.** Instruction Set Features
   ➢ Native Instructions for IEEE 754
   ➢ Byte-Addressable (8-, 16-, 32-Bit Data)
   ➢ 8-Bit Overflow Protection
**d.** 256K-Byte L2 unified Memory RAM\Cache.

### III. QUICK START WITH NON-REAL TIME PROGRAMMING

A. **General steps for non-real time implementations**:
**Step 1**. Launch code composer studio (ccsv5/6). Then go to File → New → New CCS project.
**Step 2**. Enter the details as shown in the figure below (given any name for project)

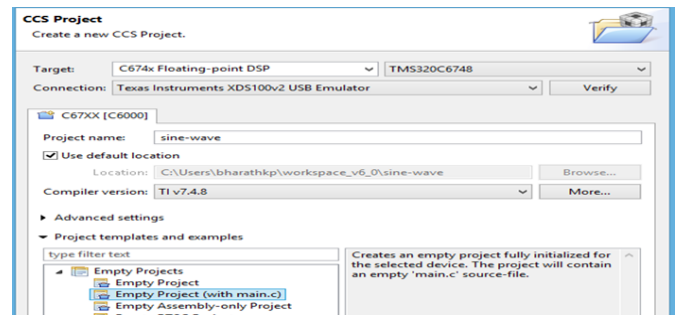

Then click → Finish.
**Step 3**. Now copy and paste the program in the source file ( main.c) and save it.
**Step 4**. Go to Project →Build project. (after build project, ***Build Finished*** message will comes in the console window as shown below

```
**** Build Finished ****
```

**Step 5**. Run →Debug.
**Step 6**. After the program is loaded successfully, click Run→Resume.

**Step 7**. To plot the graph go to: View →Expressions (now enter the output variable by clicking Add new expressions, as shown below

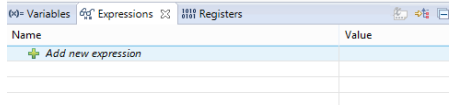

**Step 8**. Add the output variable in the expressions. right click on that variable and click the option graph.

### B. Non-real time implementations:

**Sine wave generation:** Fig.4 consist the program for simple sine wave generation. This program will generate the sine wave with 100 samples. The outputs of the sine wave generated samples are stored in the variable y. the generated sine wave is plotted as shown in fig.5.

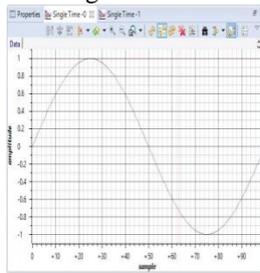

Fig.4 Program for sine wave.   Fig.5 Graph for sine wave

**Low Pass Filter and High Pass Filter:** For both Low Pass Filter (LPF) and High Pass Filter the order is chosen as 20 and the cutoff frequency is 50 Hz. The programs for both LPF and HPF are shown in fig.6 and fig.7 respectively, and their respective graphs are shown in fig.8 and fig.9.

Fig.6 program LPF.    Fig.7 Program for HPF

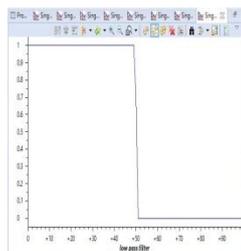 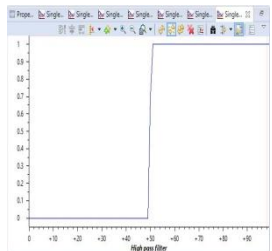

Fig.8: Graph for LPF    Fig.9: Graph for HPF

## IV. QUICK START WITH REAL TIME PROGRAMMING

Some of the real time implementation prerequisites are C6748LCDK, PC with Code Composer Studio, CRO, Audio Source, Speakers, headphone and Signal Generator.

### A. General steps for real time implementations:

**Step 1**. Create project: Goto File →new→CCS project and enter the details as done in Non-real time but only change is in "Advanced Settings" choose the Linker command file: as "**linker_dsp.cmd**"

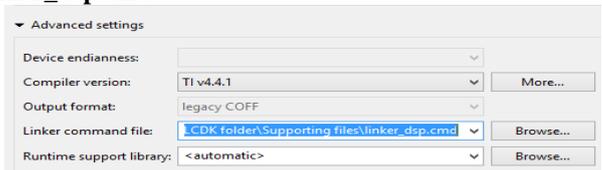

**Step 2**. Create source file to write code. In source file type (or copy) the following code.

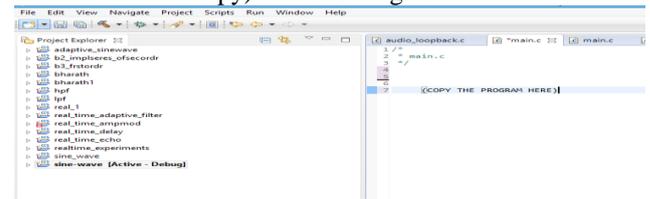

Once the program is written, save it.
**Step 3**. Now add the following library files and supporting files of codec to your project.
To add files goto Project→ Add files.

Follow the path given below to add the required files:
1. Supporting files\**evmomapl138_bsl.lib**
2. Supporting files\**L138_LCDK_aic3106_init.h**
3. Supporting files\**L138_LCDK_aic3106_init.c**
4. Supporting files\**vectors_intr.asm**

**Step 4**. Right click on the project name and choose "Show Build Settings.." → Build→ C6000 Compiler→ Include Options →Click "Add" as shown in the figure.

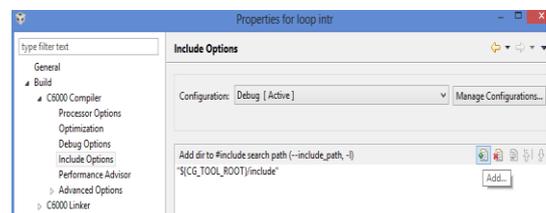

File system → C6748DSK\LCDK folder\Supporting files→bsl→Inc→OK.

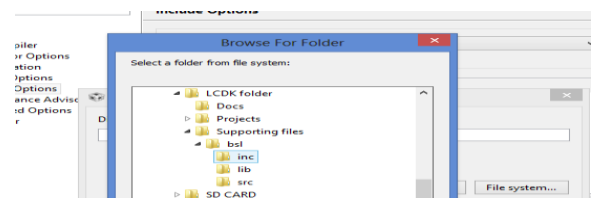

**Step 5.** Build Project, Load and Resume the project.

## B. Audio based real time implementations:

Basically, there are 3 different input/output methods are available: Interrupt method, Polling method, and DMA based method. But in this paper we adopted only interrupt method. Some of the sample and simple programs is as follows. The 48000 Hz sampling frequency, 0 dB attenuation gain is used for both ADC and DAC are used for all the real time programs.

**Audio loop back:** Audio loop back refers to the routing or transmitting the audio signal from one device to another device without any modification or processing. In this example we are playing the audio signal in PC and it passes to LCDK board via 2 way audio jack cable which is intern connected to the line-in port in the board, the output from the board is taken from the line-out port that is given to the speaker or ear phone. The program for audio loop back is as shown in the fig.10. In this program we used interrupt driven input/output method.

```
#include "L138_LCDK_aic3106_init.h"
uint32_t input;
int i = 0;
interrupt void interrupt4(void) // interrupt service routine
{
 input = input_left_sample();
 output_left_sample(input);
 return;
}
int main(void)
{
 L138_initialise_intr(FS_48000_HZ,ADC_GAIN_0DB,
 DAC_ATTEN_0DB,LCDK_LINE_INPUT);
 while(1);
}
```

Fig.10 Audio loop back program

**Audio Delay:** Generally, if the audio is played and stored in buffer and after a certain amount period of time that audio will plays back this is known as audio delay. In our example, we first assigned the buffer size of 4000 this intern acts as delay to the given audio. After some amount of delay we can able to hear that audio through the speaker.

**Audio echo:** Echo of an audio is the repetition or reflection of an audio signal which arrives at the listener with some amount of delay after the original sound of an audio. In this example also we provided delay amount of 4000 buffer size with 0dB gain. The program for audio echo is as shown in the fig.12.

## V. APPLICATIONS

*Digital alarm clock using OMAP_L138:*

Here we made a digital clock which shows time, day of the week and has an alarm. Since, the processor is faster than the normal clock approximately, 20 seconds(19.2 seconds), we have introduced a delay which would synchronize the OMAP-L138 digital clock with the normal clock. In addition to displaying time, we have added the days of the week wherein 23:59:59 would be one day and 00:00:00 would become the next day. We have also made a customized message alarm clock which would notify us at the time we specify. All we have to do before running the code is to set the hour, minute, second, day of the week(in terms of number where Sunday=1, Monday=2,...) and alarm time, and we are good to go. Fig.13 shows the flow chart of digital alarm clock.

```
#include "L138_LCDK_aic3106_init.h"
#define BUF_SIZE 400
uint16_t input,output,delayed;
uint16_t buffer[BUF_SIZE];
int i = 0; //you can vary "i"
interrupt void interrupt4(void) // interrupt service routine
{
 input = input_sample();
 delayed = buffer[i];
 output = delayed + input;
 buffer[i] = input;
 i = (i+1)%BUF_SIZE;
 output_sample(output);
 return;
}
int main(void)
{
 int i;

 for (i=0 ; i<BUF_SIZE ; i++)
 {
  buffer[i] = 0;
 }
 L138_initialise_intr(FS_8000_HZ,ADC_GAIN_0DB,
 DAC_ATTEN_0DB,LCDK_MIC_INPUT);
 while(1);
}
```

Fig.11: Audio delay program

```
#include "L138_LCDK_aic3106_init.h"
#define GAIN 0.6
#define BUF_SIZE 4000
int16_t input,output,delayed;
int16_t buffer[BUF_SIZE];
int i = 0;
interrupt void interrupt4(void) // interrupt service routine
{
 input = input_sample();  delayed = buffer[i];
 output = delayed + input;
 buffer[i] = input + delayed*GAIN;
 i = (i+1)%BUF_SIZE;
 output_sample(output);
 return;
}
int main(void)
{
 int i;
 for (i=0 ; i<BUF_SIZE ; i++)
 {
  buffer[i] = 0;
 }
 L138_initialise_intr(FS_8000_HZ,ADC_GAIN_0DB,
 DAC_ATTEN_0DB,LCDK_MIC_INPUT);
 while(1);
}
```

Fig.12: Audio echo program.

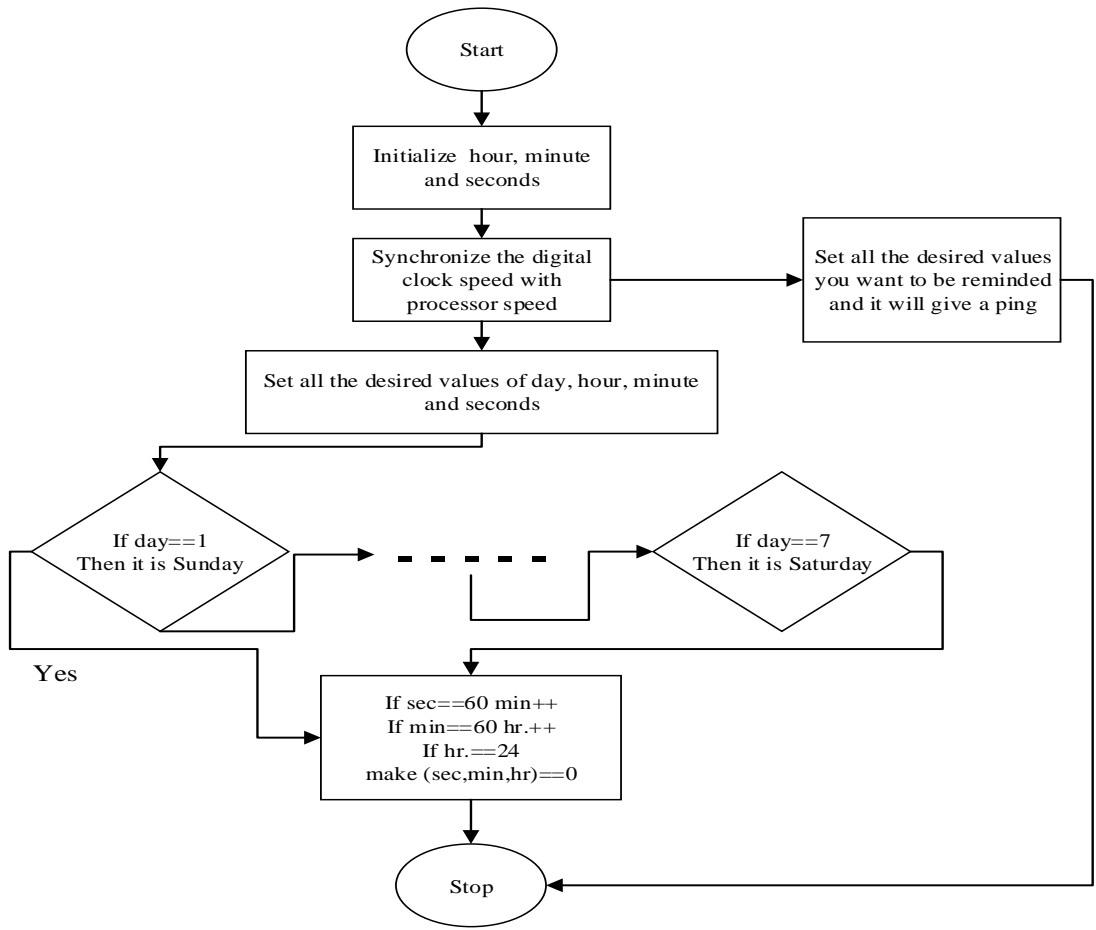

Fig.13: Flow diagram of Digital alarm clock

## VI. CONCLUSION

In this paper, we represent the basic introduction to TMS320C6748 processor which is also called as OMAP-L138 LCDK and procedural steps for both Non-Real time and Real time implementations with a set of examples like sine wave generation, LPF and HPF and their graphs under non-real time, whereas audio loop back, delay and echo is explained under real time implementations. We adopted interrupt method for all the real time programs instead of polling method. In future we will focus more real time implementations based on audio and image analysis techniques.